\documentclass[12pt]{iopart}

\usepackage{iopams}  
\usepackage{graphicx}
\usepackage{array}
\usepackage{bm}
\usepackage[english,activeacute]{babel}

\begin{document}
\pagestyle{plain}
\selectlanguage{english}

\title[Discarded particles in shower simulations effect on energy deposit]{Discarded low energy particles in extensive air shower simulations: Effect on the shower Energy Deposit}

\author{Mat\'ias Tueros}
\address{IFLP - CCT La Plata - CONICET, Casilla de Correo 727 (1900) La Plata, Argentina}
\address{Depto. de Fisica, Fac. de Cs. Ex., Universidad Nacional de La Plata,  Casilla de Coreo 67 (1900) La Plata, Argentina}
\ead{tueros@fisica.unlp.edu.ar}

\begin{abstract}
The simulation of particle cascades initiated in the atmosphere by ultra high energy cosmic ray particles involves the generation and propagation of a huge amount of particles. As it is unpractical to follow every particle to its end, particles below a certain energy ($E_{Cut}$) are discarded from the simulation.
In this article we study in detail the influence that this cut has on the total energy deposited in the atmosphere by the particle cascade in AIRES simulations. The energy deposit is directly related to the emission of fluorescence light and is critical for the accurate simulation of shower signals in fluorescence detectors.
Not correcting for the discarded particles introduces a bias on several shower observables related to the energy deposit that can range from 3 to 30\% or more depending on the $E_{Cut}$ value used. A prescription for the correct treatment of these particles is proposed, and the resulting corrections to the total energy deposit are addressed, including a new universal parametrization of the mean energy deposit per particle. The low energy cut is introduced in the simulations to reduce the required CPU time per shower at the expense of simulation accuracy. We find that a 0.4 MeV cut for electrons and 0.9 MeV cut for gammas is an adequate compromise, and that the proposed prescription is capablable of removing the bias introduced by this cut. The prescription is independent of the energy cut value and can be used to correct and compare simulations made with different energy cuts.
\end{abstract}
\vspace{2pc}
\submitto{\JPG}
\maketitle
\section{Introduction}
\label{sec:Intro}

When a Ultra High Energy Cosmic Ray (UHECR) hits the atmosphere a cascade of particles is generated. An important fraction of its energy is deposited in the atmosphere as ionization of the air molecules and atoms. A tiny, known fraction of the total deposited energy is re emitted during the de-exitation of the ionized molecules as fluorescence light that can be detected by ground telescopes.
This phenomenon is the basis for the cosmic ray fluorescence detection technique pioneered by the Fly's Eye experiment \cite{Baltrusaitis} and implemented in Hi-Res \cite{Abu} and the Auger Observatory \cite{Engeniering} to determine the energy of cosmic rays.

In this technique, one of the methods to determine the primary energy \cite{Song} is to convert the light emitted by the shower to the longitudinal profile of charged particles in the shower (using the photon yield per particle \cite{Kakimoto} \cite{Nagano}) and then estimate the calorimetric energy integrating the charged particle profile using the mean ionization loss rate, obtained from air shower simulations. The estimation of the energy deposited on the shower simulation is thus of central importance for the reconstruction and analysis of the measured signals.

The estimation of the aperture and exposure of the detector also relies heavily on the correct estimation of the energy deposit. For the simulation of the signal a shower would generate on the detectors, the energy deposit is used to estimate the photon emission that is then propagated to the detectors taking into account all the atmospheric effects. The amount of light arriving to the detector is then used as input to simulate the detector response.

As it is unpractical (if not unfeasible) in UHECR simulations to follow all the particles to their rest, a low energy cut is made. Particles with energy below the cut value ($E_{Cut}$) are not tracked any more by the simulation program and the fate of the energy they carry is not determined. A fraction of this energy might end deposited by ionization in the atmosphere and would thus contribute to the energy deposit. This introduces a bias on all the observables related to the energy, like the total energy deposit, the total electromagnetic energy and the mean energy deposit per particle, parameters usually used as input parameters in the simulation and reconstruction of UHECR showers signals. The introduced bias has a dependency with $E_{Cut}$ value, making comparison of simulations made with different cuts difficult.

In this article we present an algorithm to estimate and correct for the bias introduced by the discarded low energy particles in AIRES simulations \cite{AiresManual}, following a prescription similar to the one presented in \cite{Barbosa} and \cite{Risse} for CORSIKA simulations. The algorithm is tested on a $2\times10^{4}$ shower library described in section \ref{sec:Simulations}, paying special attention to the effect this correction has on the shower energy deposit and the mean energy loss per particle. To this end, in section \ref{sec:EnergyBalance} we make a careful study of the energy balance in the simulation and show the relevance of the discarded low energy particles and in section \ref{sec:EDep} we show how to include the contribution of these particles to the shower energy deposit. In section \ref{sec:MeanEdep} we show the effect this has on the average energy deposit per particle, and present a corrected universal parametrization with shower age.

The low energy cut is introduced in the simulations to reduce the required CPU time per shower. Increasing $E_{Cut}$ value reduces the amount of CPU time per shower, at the expense of increasing the total amount of energy discarded from the simulation, making the necessary correction more important. We address the influence of $E_{Cut}$ value on the CPU time and on the amount of discarded energy in section \ref{sec:InfluenceEnergyThreshold}.

\section{About the simulations}
\label{sec:Simulations}

The quantitative results presented in this work are based on a particular but representative set of AIRES simulations.The AIRES code has been extensively used by many scientist around the world for the past ten years, and has become one of the standard simulation codes in the field.

AIRES includes the most important processes that may undergo shower particles from a probabilistic point of view. For the estimation of the shower energy deposit, electrodynamical processes play a central role. Pair production,electron-positron annihilation, bremsstrahlung (in electrons, positrons and muons), muonic pair production, knock-on electrons ($\delta$ rays), Compton effect, photoelectric effect, Landau-Pomeranchuk-Migdal (LPM) effect and dielectric suppression effect are all taken into account.

Collision and bremsstrahlung effects have a low energy divergence and can not be simulated to zero energy, since the amount of emitted particles would also diverge (not to mention the complexity of simulating, for example, electron capture by an atom or air molecule). In all simulation codes a threshold energy is imposed, below which emission due to these processes is not longer simulated. In AIRES the emission of $\delta$ rays and bremsstrahlung radiation is limited to 1 MeV and 0.1 MeV respectively. The particle energy loss due to the emission below these thresholds is treated with the continuous energy loss approximation during particle propagation, using a parametrization taken from GEANT3 simulations. 

These \textit{emission} energy cuts are independent of and must not be confused with the \textit{simulation} energy cuts studied in this article, that set the energy below which particles are \textit{discarded} from the simulation. In AIRES, different $E_{Cut}$ values can be set for different particle species. The ones more important for the energy deposit study made in this work are electron/positron and gammas $E_{Cut}$ values.

The shower library used in this work was generated at the in2p3 computing center \cite{Lyon} and has the following characteristics:\\
\begin{small}
Series name: AMgeLyonExtDvezpShb, generated by Sergio Sciutto\\
Hadronic models: QGSJET-II and SIBYLL\\
Primary particles: Proton, Iron and Photon (With MAGICS Preshower)\\
$Log_{10}(Energy)$: 17.5, 18, 18.5, 19, 19.25, 19.5, 19.75, 20, 20.25, 20.5\\
Zenith (deg):0, 18, 25, 32, 36, 41, 45, 49, 53, 57, 60, 63, 66, 70, 72, 75, 78, 81, 84, 87\\
Thinning Energy Rel.: $1.0\times10^{-6}$     Thinning W. Factor: 0.1 \\
Gamma Cut Energy = 0.9 MeV\\
Electron/Positron Cut Energy = 0.4 MeV\\
Muon Cut Energy = 2.5 MeV\\
Meson Cut Energy = 4.5 MeV\\
Nucleons Cut Energy = 95 MeV\\
Thinning refers to a statistical sampling procedure to reduce computing time by simulating only a representative sample of all the particles. Thinning and thinning related parameters are explained in detail in the AIRES user manual \cite{AiresManual}.\\
\end{small}

Note that $E_{Cut}$ for gamma and electron/positron are low enough to determine unambiguously the fate of the particles. Gamma below 900KeV will not be able to generate pairs, and will deposit all their energy. Electrons generated will also deposit all their energy until captured, and positrons will annihilate producing 2 or 3 gammas with energy below 1 MeV and thus unable to generate new pairs.

The results presented in this work exclude photon primaries to make figures clearer and  keep the discussion simple. The algorithms presented in this article are perfectly applicable to photon (or electron) primaries but purely electromagnetic showers have qualitative and specially quantitative differences with hadronic showers that make it cumbersome to present all the results together. An article focusing in these differences is in preparation.

\section{Shower Energy Balance}
\label{sec:EnergyBalance}
During the evolution of a UHECR shower the cosmic ray primary energy is distributed among all the particles on the cascade. Part of this energy is deposited in the atmosphere, part is lost to the neutrino channel and part arrives at ground level as kinetic and rest mass energy of the particles. In simulations, a part is also carried away by the particles that fall below the low energy cut and are removed from simulation.\\
To check the AIRES simulations and the treatment of the low energy particles in this work, we devote this section to the energy balance in the shower.\\
In AIRES, the following information is readily available:\\
$E_{Dep}$: Total energy \textbf{deposited} on the atmosphere by all particles in the simulation (AIRES table 7993).\\
$E_{Dis}$: Total kinetic energy of all particles \textbf{discarded} by the low energy cut (AIRES table 7793).\\
$E_{\nu}$: Total Energy lost to the \textbf{neutrino} emission channel (AIRES .sry file).\\
$E_{Ground}$: Total kinetic energy of all particles \textbf{reaching ground level} (AIRES table 1793).\\
\setcounter {figure} {0}
\begin{figure} 
\begin{center} 
\includegraphics[angle=0,scale=0.8]{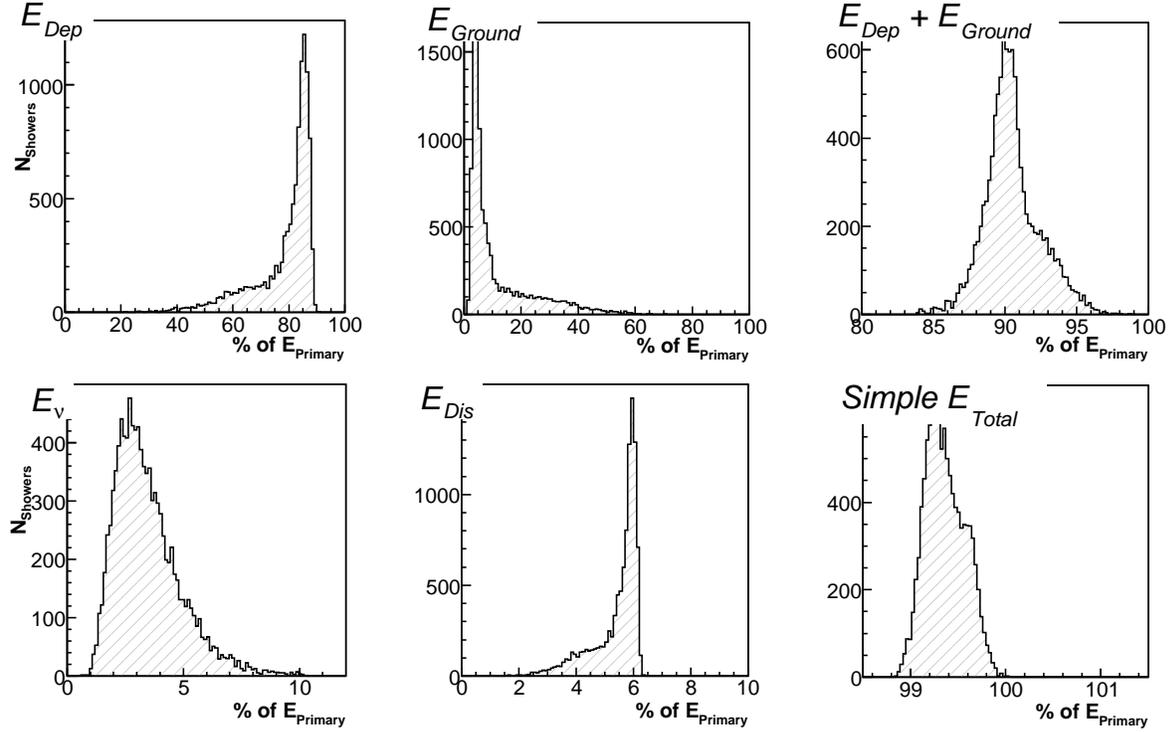} 
\caption{Histograms showing the contributions to the total energy in the shower for our simulation sample.
$E_{Dep}$:Total energy deposited on the atmosphere by all particles. $E_{Ground}$: Total kinetic energy of all particles reaching ground level. $E_{\nu}$: Total Energy lost to the neutrino emission channel. $E_{Dis}$:Total kinetic energy of all particles discarded by the low energy cut. $Simple E_{Total}$ is the sum $E_{Dep} + E_{Dis} + E_{\nu} + E_{Ground}$ and constitutes the simplest way of computing the total energy on the shower.
\label{gr:EnergyBalanceContributions}
}
\end{center}
\end{figure}

With this information, we can calculate the total energy of the shower as

\begin{equation} \label{eq:simpleEtotal}
Simple E_{Total}= E_{Dep} + E_{Dis} + E_{\nu} + E_{Ground}
\end{equation}

The sum of these four components is the simplest energy balance of the simulation that can be made. Figure \ref{gr:EnergyBalanceContributions} shows the distribution of the contribution of the different terms of \ref{eq:simpleEtotal} in our simulation sample. $E_{Dep}$ and $E_{Ground}$ have a strong dependence with the shower zenith, since inclined showers traverse a longer distance in the atmosphere and deposit more energy, producing the long tails in the distributions. The sum of these two components account for 90$\pm$5 \% of the primary energy.

Neutrino energy ($E_{\nu}$) accounts for an average 3.4 \% of the primary energy in our sample, and also have a strong dependence with the shower zenith. In extreme cases it can reach up to 10 \%.

Figure \ref{gr:EnergyBalanceContributions} also shows that particles discarded by the low energy cut, usually ignored, can have an important contribution to the total energy. Even using a low $E_{Cut}$ value like the one used in our sample simulations, they represent an average 5.5 \% off the total energy. 

The four terms in \ref{eq:simpleEtotal} do not account for all the primary energy but for 99.35 $\pm$ 0.4 \% of it . Some showers are still missing about 1\% and there is an overall 0,65 \% negative bias. This unaccounted energy is carried away by the rest mass of the discarded low energy particles, that was not included in the $E_{Dis}$ term of \ref{eq:simpleEtotal}.

A correction for the rest mass of low energy particles is not straightforward. In AIRES not all particle information is saved in the simulation. Low energy particles are classified in "gammas" (table 7001), "electrons"(table 7005),"positrons" (table 7006),"muons +" (table 7007),"muons -" (table 7008), "other charged" (table 7291) and "other neutral" (table 7292). 

\begin{figure} 
\begin{center} 
\includegraphics[angle=0,scale=0.8]{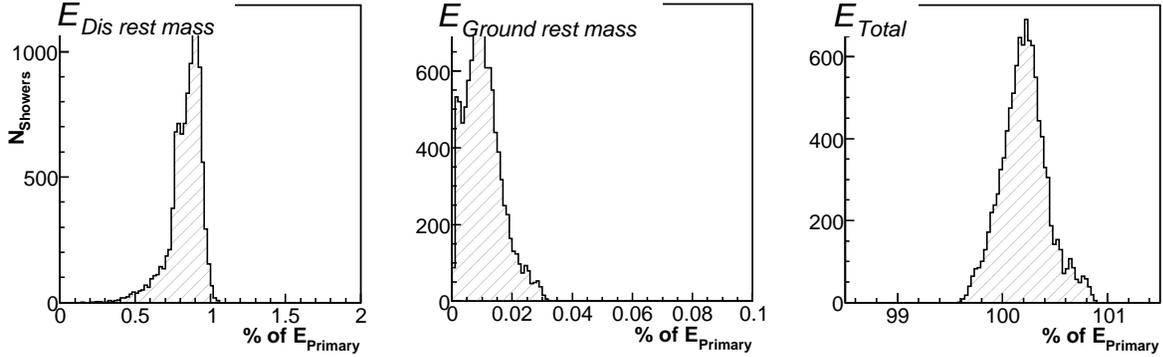} 
\caption{Histograms showing the contribution of the rest mass of discarded particles and ground particles to the total energy in the shower for our simulation sample. $E_{Dis\:rest\:mass}$:Total rest mass energy of all the discarded particles. $E_{Ground\:rest\:mass}$: Total rest mass energy of all particles reaching ground level. $E_{Total}$: Total Energy, calculated as $Simple E_{Total}$ + $E_{Dis\:rest\:mass}$ + $E_{Ground\:rest\:mass}$.}
\label{gr:EnergyBalanceContributions2}
\end{center}
\end{figure}

The particles in "other charged" and "other neutral" are assumed to be pions, the lightest unclsassified particle. The contribution to the total energy from this miscellaneous groups is of the order of 0.4\% when we choose this mass. This is a lower limit, as some of these particles could be nuclei,nucleons or kaons. The distribution of this correction can be seen in figure \ref{gr:EnergyBalanceContributions2} labeled as $E_{Dis\:rest mass}$.

Not all low energy particles are "created" at the expense of primary energy. Most low energy electrons are taken from the atmosphere in Compton or knock on collisions. The rest mass of low energy electrons is discarded as a first approximation.

For completeness, ground particles rest mass will also be included although it was found to contribute in average with less than 0,01\% to the primary energy. Ground particle classification in AIRES is more complete, including classes for protons, neutrons, pbar, charged pions and charged kaons (tables 1001 to 1293). Problematic classifications are again "other neutral", "other charged" and "nuclei" (all empty in our simulation sample). An average rest mass of 1 GeV is suggested for this last three classes, providing a lower limit. As in the low energy particles case, ground electrons rest mass is not included. The distribution can be seen in figure \ref{gr:EnergyBalanceContributions2} labeled as  $E_{Ground\:rest\:mass}$.

With the introduction of the rest mass correction the total energy can now be written as

\begin{equation} \label{eq:Etotal}
E_{Total}= E_{Dep} + E_{Dis} + E_{\nu} + E_{Ground} + E_{Dis\:rest\:mass} + E_{Ground\:rest\:mass}
\end{equation}

After applying this correction we get a mean total energy of 100.2 \% of the primary energy, with a standard deviation of 0.2 \% (see top right of figure \ref{gr:EnergyBalanceContributions2}). This results confirm that assuming "other particles" to be pions is a good lower limit and that a more detailed particle classification is not necessary, as this will not change the results presented in this article in an appreciable way.
The total energy of the shower is expected to be always slightly higher than the primary energy, as the medium provides a small amount of energy to the shower each time a particle breaks an air nucleus. We can see that although considering "other particles" to be pions provides a good lower limit,there are still some showers below a 100\%. A study on this small "excess energy" might be interesting from a theoretical point of view and is planned for a future article.

\section{Energy Deposit off Low Energy Particles}
\label{sec:EDep}
We have shown in the last section that low energy particles constitute an appreciable percentage of the primary energy. Since a fraction of this energy can be deposited in the atmosphere, the regular energy deposit estimation in the shower simulation must be corrected.

AIRES provides tables with the longitudinal development of several observables, among them the total energy deposit (table 7993) and the energy of the particles discarded by the low energy cut ($E_{Dis\:x}$), discriminated by particle type (tables  7501 to 7892).

Not all the particles lost due to the low energy cut will deposit all their energy, and care must be taken when correcting the total energy deposit. Neutrinos from the decay of the discarded low energy muons, for example, will not deposit their energy in the atmosphere. Discarded low energy positrons, on the other hand, will surely annihilate with an electron on the air, giving gammas that will deposit almost all their energy. Low energy Pions and other Hadrons might undergo even more complex interactions, with an a priori unknown amount of deposited and missing energy.

An accurate enough correction can be made by adding the contribution from electrons, gammas, muons and other particles (assumed to be pions) using the results from \cite{Barbosa} in which the fraction of "releasable" energy was calculated for each particle species using Geant4 simulations: 0.997 for gammas and neutral pions, 0.998 for e+/e-, 0.425 for muons and 0.46 for charged pions. The prescription suggested to calculate the amount energy deposited by the discarded particles ($E_{Dis}^{Deposit}$) is

\begin{eqnarray} \label{eq:EDisDeposit}
E_{Dis}^{Deposit} & = & 0.997E_{Dis\:\gamma}+0.997E_{Dis\:\pi^{0}}+0.998E_{Dis\:e^{+/-}}+\nonumber\\
                  &   & 0.425E_{Dis\:\mu^{+/-}}+0.46E_{Dis\:\pi^{+/-}}
\end{eqnarray}

\begin{figure} 
\begin{center} 
\includegraphics[angle=0,scale=0.8]{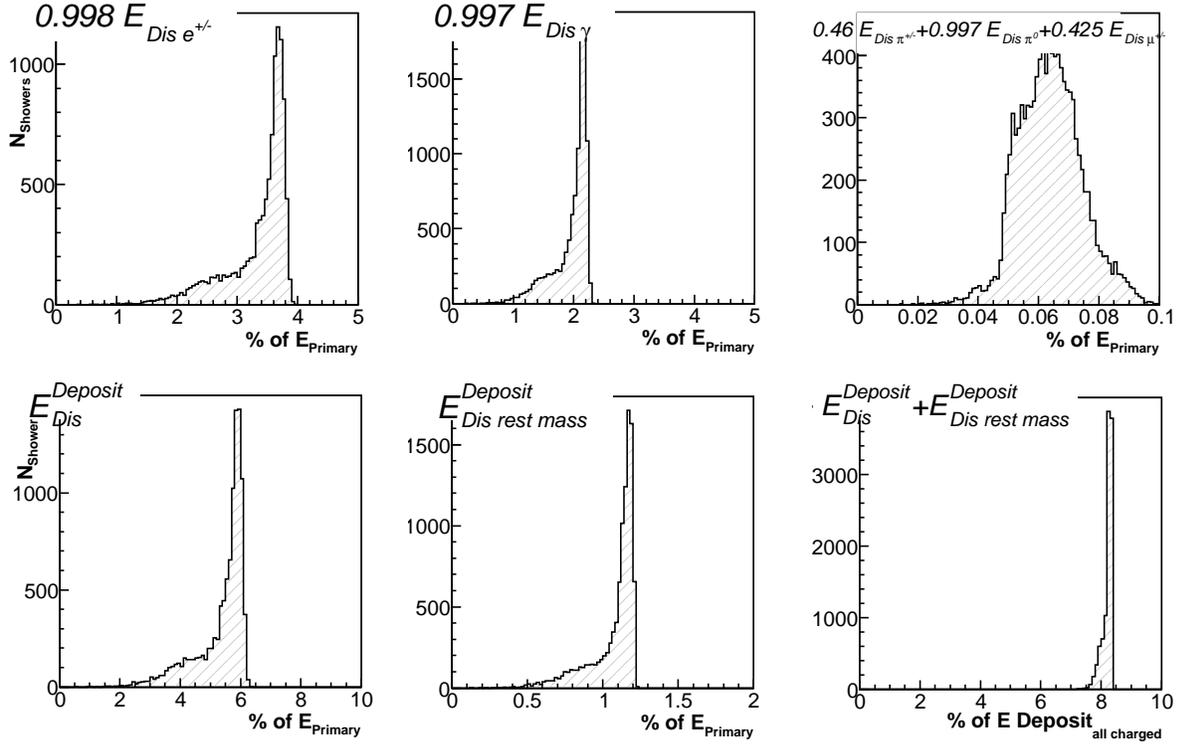} 
\caption{Histograms showing the contribution of the discarded particles to the total energy deposit in the shower for our simulation sample. The top row shows the different terms of \ref{eq:EDisDeposit}. The bottom row shows the contribution of the kinetic energy and the rest mass of the discarded particles and the \% correction these represents to the total energy deposit of all the charged particles.
\label{gr:CalLowE}
}
\end{center}
\end{figure}

The distribution of this correction for our simulation sample and the contribution from each term in \ref{eq:EDisDeposit} can be seen in figure \ref{gr:CalLowE}. It can be seen that the contribution from muons and pions is negligible when compared to electrons and gammas. More than 98 \% of the low energy particles are electrons and gammas, that will in the end deposit almost all of their energy, making the results insensitive to the details of the contribution from other particles. For the $E_{Cut}$ values used in our simulations, low energy particles deposit on average 5.4\% of the primary particle energy.

A second order correction can be made including the contribution of the rest mass of discarded unstable particles. Unstable particles that where discarded will eventually decay or annihilate, giving electromagnetic particles that will deposit their energy producing ionization and more fluorescence light. At this level of detail, the contribution of energy from the medium (the air) that is in fact "releasable" energy must also be considered. In the annihilation processes, the rest mass of the antiparticle is added to the shower energy pool, and part of that energy is also capable of producing ionization.

Being this a small correction, and following the suggestion in \cite{Risse} the "depositable" energy was assumed to be 1/3 of the rest mass for particles more likely to decay (muons and pions) and twice the rest mas for positrons, considered to annihilate and produce gammas that will deposit all their energy. As stated previously "other" particles are assumed to be pions. The suggested prescription to treat the energy deposit from the rest mass of the discarded particles ($E_{Dis\:rest\:mass}^{Deposit}$) is

\begin{equation} \label{eq:EDisRestMass}
E_{Dis\:rest\:mass}^{Deposit} =\frac{1}{3}m_{\mu}N_{Dis_{\mu}}+\frac{1}{3}m_{\pi}N_{Dis_{\pi} }+2m_{e}N_{Dis_{e^{+}}}
\end{equation}

The number of discarded low energy particles ($N_{Dis_{x}}$) is available in AIRES tables 7001 through 7293. In the simulations made for this work, the correction due to the low energy particles rest mass is around 1\% of the primary energy, and is almost entirely due to the annihilation of low energy positrons (See figure \ref{gr:CalLowE}. 

The total correction due to the low energy particles is obviously dependent on $E_{Cut}$, and this will be addressed in section \ref{sec:InfluenceEnergyThreshold}. For the simulations made on this work, it has a mean value of about a 7.85 $\pm$ 0.33 \% of the total energy deposit and is almost independent of the shower details. When compared to the energy deposited by the all the charged particles above $E_{Cut}$ (the quantity usually used for shower fluorescence light simulations) this correction represents a 8.3 $\pm$ 0.3 \% increase (figure \ref{gr:CalLowE}).

%


As the energy deposit is more or less proportional to the emitted light, not correcting for the low energy particles introduces a proportional negative bias in the signal simulation producing dimmer showers that are harder to detect, reducing the detector aperture and that will be reconstructed as events of lower total energy. The estimation of the mean energy deposit per particle used in estimation of the primary energy in the reconstruction of real detector events is also proportionally affected.

\section{Mean Energy deposit per particle}
\label{sec:MeanEdep}
The determination of the primary particle energy using the fluorescence technique is considered a calorimetric measurement, in which the atmosphere acts as a calorimeter where most of the primary particle energy is deposited through ionization.

The usual procedure \cite{Abu} \cite{Song}, consists in converting the measured light profile $N_{ph}(X)$ in a region $\Delta X$ to a number of charged particles profile $N_{ch}(X)$ 

\begin{equation} \label{eq:Nch}
N_{ch}(X)=\frac{N_{ph}(X)}{\Delta X\times{}Y}\frac{1}{g_{geo}\times{}g_{det}\times{}g_{atm}}
\end{equation}

using the photon yield per electron per $g/cm^2$ of traversed air $Y$ and the corresponding geometrical, detector and atmospheric related correction factors .

This profile is integrated to get the total number of charged particles and multiplied by the mean energy deposit per particle per $g/cm^2$ $<\alpha_{eff}>$ 

\begin{equation} \label{eq:Eem}
E_{em}=<\alpha_{eff}>\int{N_{ch}(X)dX} 
\end{equation}

To estimate $<\alpha_{eff}>$, the weighted average of the mean energy deposit per particle must be calculated using Monte Carlo simulations, this is

\begin{equation} \label{eq:alpha}
<\alpha_{eff}>=\frac{\displaystyle\sum_{i=1}^{Ground} \frac{E_{Dep}(X_{i})}{N_{ch}(X_{i})\Delta X_{i}}N_{ch}(X_{i})}{\displaystyle\sum_{i=1}^{i_{Ground}}N_{ch}(X_{i})}= \displaystyle{\frac{E_{Dep}^{Total}}{N_{ch}^{Total}\Delta X}} 
\end{equation}

Where $E_{dep}(X_{i})$ is the energy deposited in a region $\Delta X_{i}$ by the $N_{ch}(X{i})$ charged particles crossing the same region. 

Using the prescription for the energy deposit of the discarded low energy particles discussed on section \ref{sec:EDep}, the computation of $E_{Dep}^{Total}$ is straightforward, and to a certain extent independent of the simulation $E_{Cut}$ value as it will be shown in section \ref{sec:InfluenceEnergyThreshold}. 

\begin{equation} \label{eq:EDepTotal}
E_{Dep}^{Total}=E_{Dep} +  E_{Dis}^{Deposit} + E_{Dis\:rest\:mass}^{Deposit}
\end{equation}

The total number of charged particles $N^{cut}_{ch}$ is not independent of $E_{Cut}$, but it is possible to apply a correction to render it independent of the low energy cut using the parametrization of the charged particles energy spectrum at shower maximum from \cite{Nerling}, 

\begin{equation} \label{eq:N0}
N^{0}_{ch}(X{i})=N^{cut}_{ch}(X{i})/(1-0.045E_{cut}/MeV)
\end{equation}

for our simulations, the $E_{Cut}$ value for electrons is 0.4 MeV and the resulting correction factor is 0.982.

Historically, the standard way of calculating the mean energy deposit per charged particle was to divide the total energy deposited by the total number of charged particles. For our simulations, the mean energy deposit value obtained in this way is 2.195 $MeV/g/cm^2/particle$, in excellent agreement with the historical values used Hi-Res and other studies \cite{Song}.\\

If we consider the energy that would have been deposited by the discarded low energy particles $E_{Dis}^{Deposit}$ as defined on \ref{eq:EDisDeposit}, the mean value is 2.35 $MeV/g/cm^2/particle$, a 7 \% increase. Including the rest mass energy correction $E_{Dis\:rest\:mass}^{Deposit}$ as proposed on \ref{eq:EDisRestMass} the mean value is 2.375 $MeV/g/cm^2/particle$, a 8.2 \% increase. This is a correction comparable to the correction found in section \ref{sec:EDep} for the total energy deposit, as expected.\\


Note that \ref{eq:Eem} implies that any change in $<\alpha_{eff}>$ will have a direct impact on the estimation of the primary energy. Simulations made with higher $E_{Cut}$ values are expected to scale up this correction accordingly.

\begin{figure} 
\begin{center} 
\includegraphics[angle=0,scale=0.8]{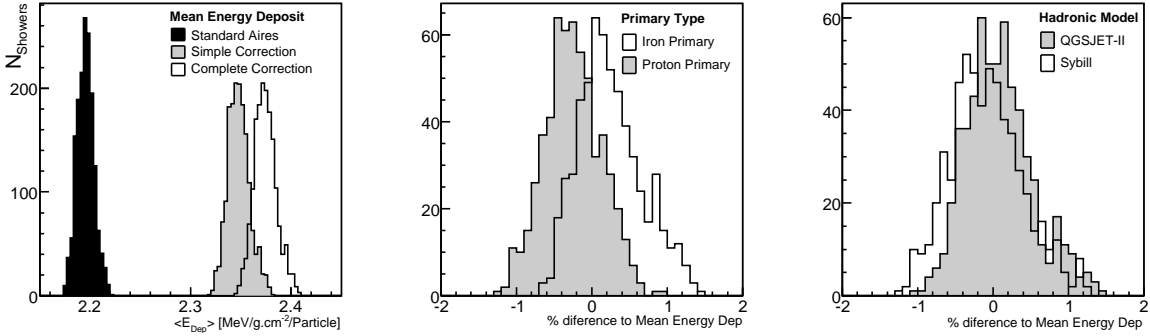}
\caption{Mean Energy Deposit per particle per $g/cm^2$ of air. Left: Values obtained with the full, the simple and without correction for the discarded low energy particles. Center: Difference to the mean with by primary particle type. Right: Difference to the mean by hadronic model used in the simulation.}
\label{gr:MeanEdep}
\end{center}
\end{figure}

The mean energy deposit is independent (within 0.5\%) of the hadronic interaction model used and of the primary type, as shown in figure \ref{gr:MeanEdep}. This comes to no surprice since the energy deposit depends mostly on the details of the low energy electromagnetic models. Note that the spread in the distribution of the mean energy deposit for our sample is only 2\%, showing little dependence with the other shower parameters .

\begin{figure} 
\begin{center} 
\includegraphics[angle=0,scale=0.8]{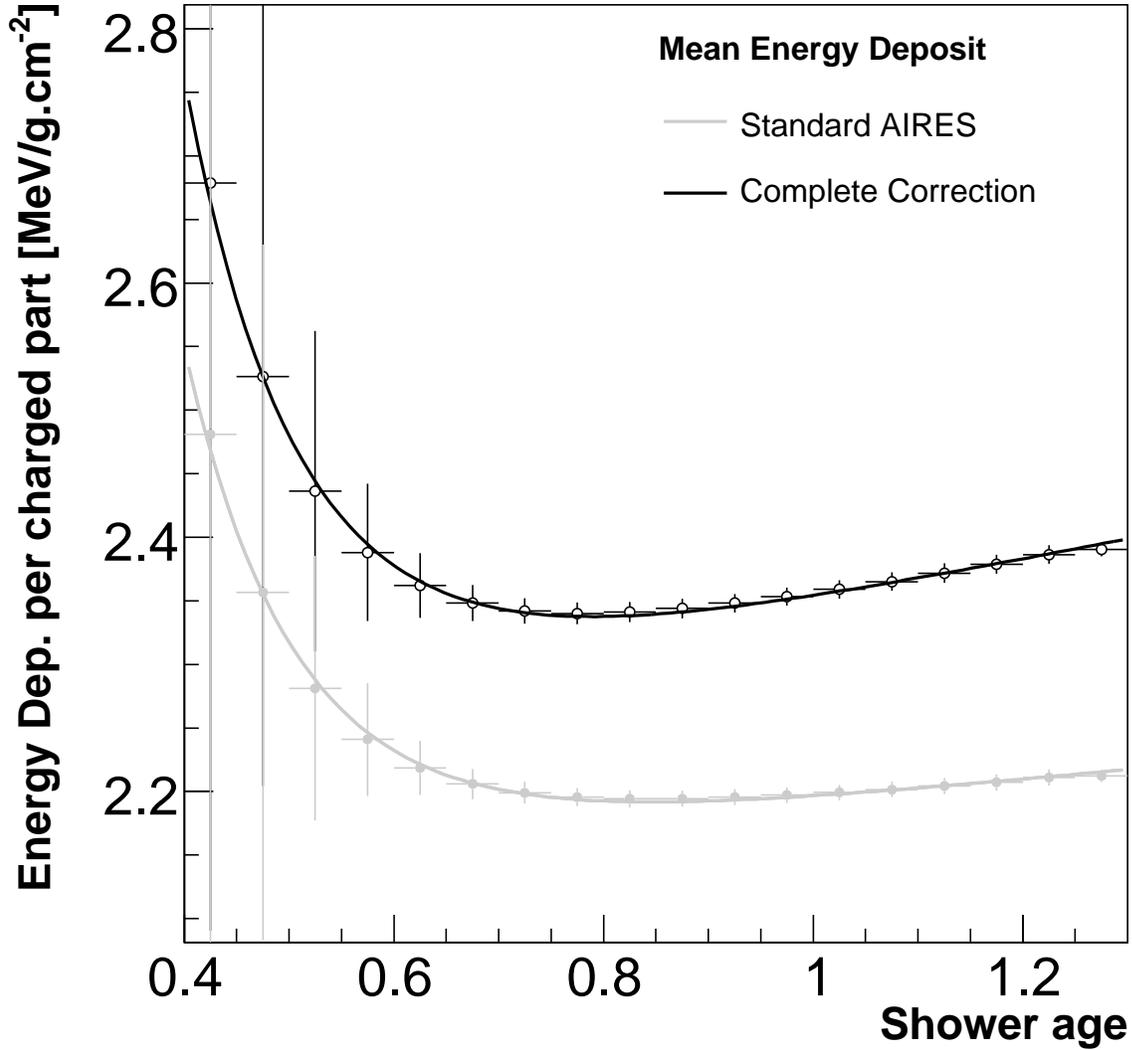}
\caption{Mean Energy Deposit per particle per $g/cm^2$ of air vs Shower Age Values obtained with and without correction for the discarded low energy particles.}
\label{gr:MeanEdepvsAge}
\end{center}
\end{figure}

The mean energy deposit per particle has some dependence with the stage of shower development (the shower age, see \ref{eq:age}), as can be seen in figure \ref{gr:MeanEdepvsAge}. This curve and the fluorescence yield can be used to convert a \emph{Fluorescence photons} longitudinal profile to a \emph{Number of Charged Particles} profile, using \ref{eq:Nch}

When the shower is young and the number of particles is still relatively low, shower to shower fluctuations dominate the dispersion in the energy deposit. Above age 0.7, when shower to shower fluctuations are no longer an issue, this curve can be considered universal within 1\% since the mean energy deposit per particle is virtually independent of the primary particle, its energy, the shower geometry and the hadronic models considered in this article. A parametrization of this curve can be achieved using the function


\begin{equation}
       <\alpha_{eff}(\mathbf{age})>=\frac{E_{Dep}^{Total}(\mathbf{age})}{N_{ch}^{0}(\mathbf{age})\:\Delta X}=\frac{A}{(B + \mathbf{age})^{C}} + D + E\mathbf{age}
\end{equation}

\begin{equation} \label{eq:age}
       \mathbf{age}=3/(1+\frac{2X_{max}}{X})
\end{equation}

where A=0.9921, B=0.67, C=9.7878, D=2.1821, E=0.1656  and $X_{max}$ is the depth of the shower maximum in $g/cm^{2}$.

\section{Dependency with the Low Energy Cut value}
\label{sec:InfluenceEnergyThreshold}
The algorithms presented in this article to estimate the contribution of the particles discarded by the low energy cut are independent of the cut value, and have been tested up to 100 MeV.

\begin{figure} 
\begin{center} 
\includegraphics[angle=0,scale=0.8]{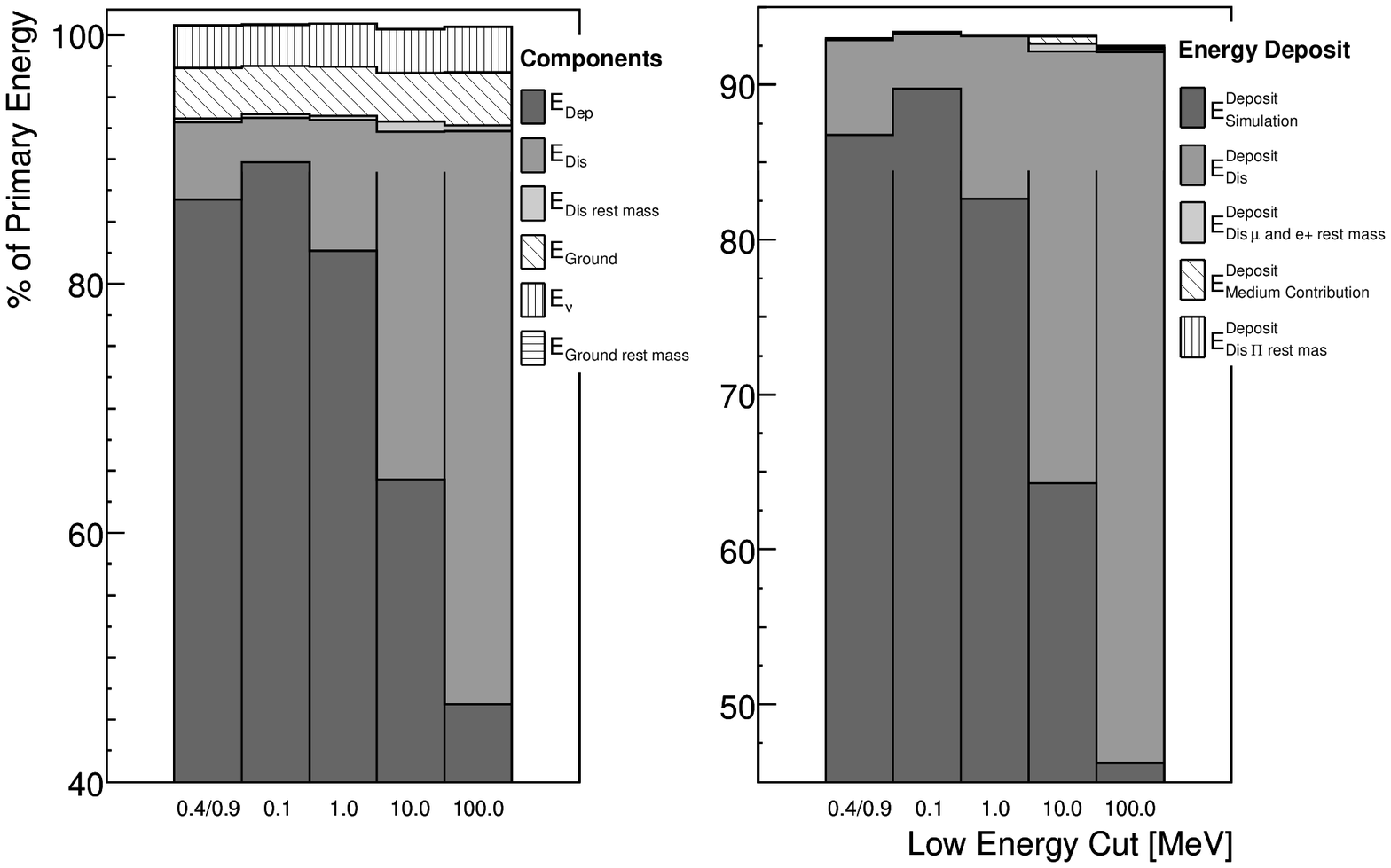} 
\caption{Components of \ref{eq:Etotal} for the total energy (left) and of \ref{eq:EDepTotal} for the energy deposit (right) of 1 EeV,60 deg zenith proton showers for different values of $E_{Cut}$. The total energy remains constant within shower to shower fluctuations, while deposited energy is traded for discarded particles energy as $E_{Cut}$ rises. The total energy deposit is also constant when the energy deposit in the simulation is corrected for the discarded low energy particles kinetic and rest mass energy deposit}
\label{gr:Ecut}
\end{center}
\end{figure}
 
The higher the $E_{Cut}$ value the higher the amount of discarded energy, and the impact of the corrections presented in the previous sections on the shower observables. To illustrate this, figure \ref{gr:Ecut} (left) shows the dependence of the total amount of discarded energy for a set of 10 showers from 1 EeV, 60 deg zenith proton primaries. It can be seen that as $E_{Cut}$ is raised, the amount of discarded energy varies accordingly but the sum of the various energy terms of \ref{eq:Etotal} remains constant within shower to shower fluctuations. 

Figure \ref{gr:Ecut} (right) illustrates how the energy deposit on the simulation is lowered as more and more energy is lost to the low energy particle cut. But if we consider the energy deposit and the corrections proposed by \ref{eq:EDisDeposit} and \ref{eq:EDisRestMass}, we see that the total energy deposit remains independent of $E_{Cut}$ within shower to shower fluctuations. This provides confidence in the proposed algorithms to account for the low energy particles energy, and thus to apply them to enable comparisons between simulations made with different energy cuts.

\begin{figure} 
\begin{center} 
\includegraphics[width=10cm,height=10cm]{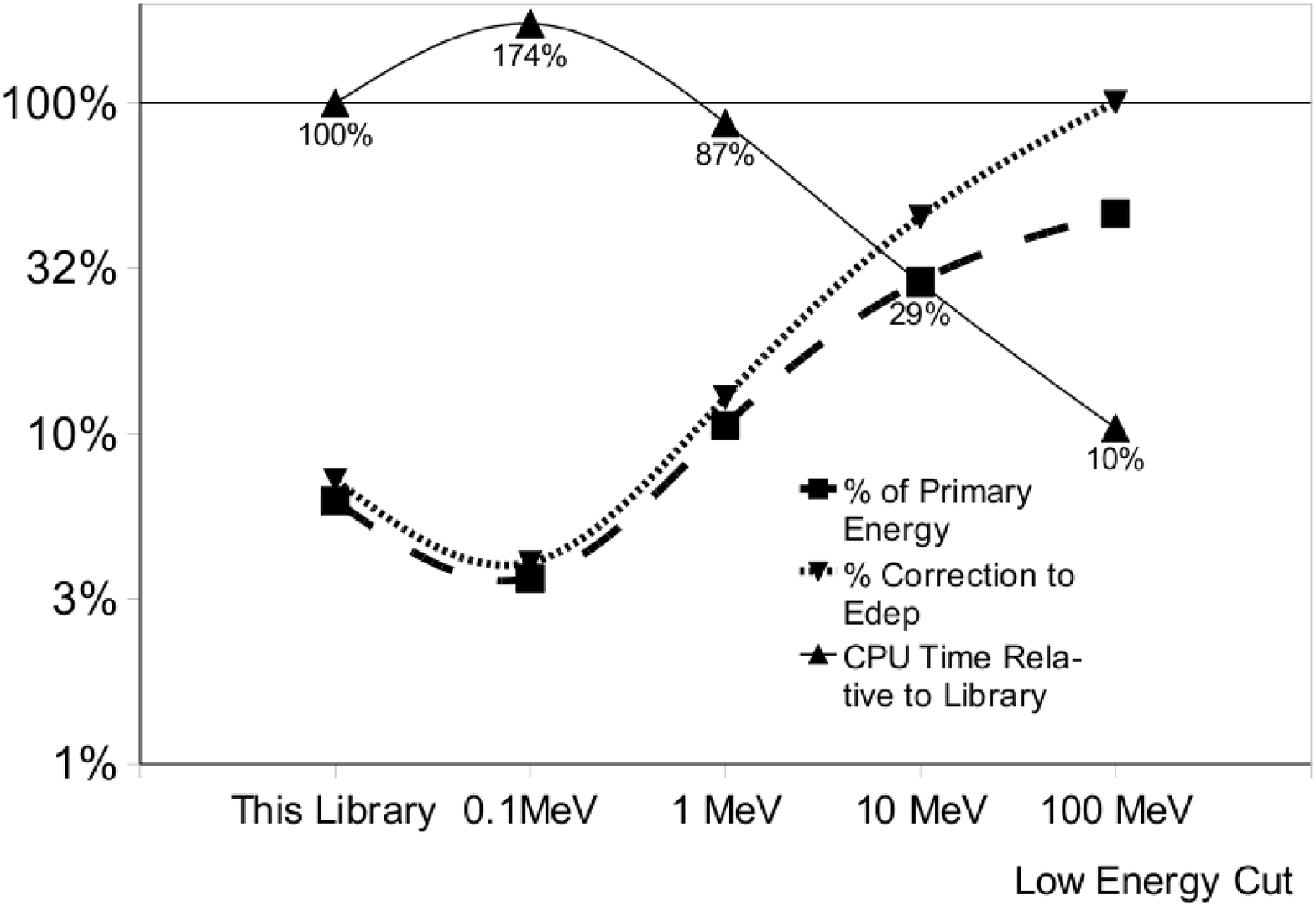}
\caption{Influence of $E_{Cut}$ value on CPU time and \% correction to the Energy Deposit. This Library has a 0.9 MeV cut for Gammas and 0.4 MeV cut for electrons and positrons}
\label{gr:CPUtime}
\end{center}
\end{figure}

For this example 1 EeV proton with the cuts set identical to the ones used in our shower library (0.4 MeV for electrons and 0.9 MeV for gammas) we see that the discarded particles account for 6.3\% of the primary energy. When all energy cuts are set at 0.1 MeV (the minimum value AIRES can handle) this goes down to 3.6\% of the primary energy. Setting the energy cut at 1 MeV it goes to 10.6\% , at 10 MeV to 29 \% and at 100 MeV to 46\% of the primary energy (Figure \ref{gr:CPUtime}). In this extreme case the low energy particles correction to the energy deposit is around a \%100 of the normal energy deposit. From this, it is clear that extreme care must be taken in the analysis of the results of simulations with energy cuts above 1 MeV when the estimation of the energy deposit is important.

Lowering the low energy cut value to minimize these corrections have a big impact on the CPU time. Going from 1 MeV to 0.1 MeV energy cut doubles the required CPU time, while lowering it to 10 MeV reduces it by nearly a factor of 3 as shown in figure \ref{gr:CPUtime}. The energy cuts used in our library are considered a good trade off, with nearly half the correction found with 1 MeV low energy cut and only 20\% more CPU time.

\section{Conclusions}
\label{sec:ConclusionsEdep}

We have shown the important influence the low energy cut has on the shower energy deposit in AIRES simulations. Not correcting for this effect was shown to introduce a negative bias on the total energy deposit that can have an important impact on shower reconstruction and shower signal simulations. The introduced bias depends on the energy cut value used in the simulation, and goes from around 3\% of the primary energy at 0.1 MeV energy cut to 30\% of the primary energy at 10 MeV energy cut on average.

An algorithm to correct for this bias independently of the energy cut value used in the simulations was presented and tested for $E_{Cut}$ between 0.1 MeV and 100 MeV successfully, making now possible to compare energy deposit results from simulations made with different $E_{Cut}$ values. Using this algorithm on a large set of simulations we computed and provided a new universal parametrization of the mean energy deposit per particle with the shower age, that can be used for the reconstruction of the primary energy of UHECR shower detected with the fluorescence technique.

Finally we studied the dependence CPU time has with $E_{Cut}$ value and found that a 0.4 MeV cut for electrons and 0.9 MeV for gammas needs only an average 6.3\% correction due to the discarded low energy particles, giving an adequate trade-off between precision in the energy deposit determination and the required CPU time.

\ack
We thank H. Wahlberg for his help running the analysis scripts in Lyon, and A. Mariazzi and S. Sciutto for usefull comments and proof reading the manuscript. M. T. is supported by a grant of the Consejo Nacional de Investigaciones Científicas y Tecnológicas de la República Argentina.

\end{document}